\documentclass[prl,twocolumn,showpacs,superscriptaddress,preprintnumbers,amsmath,a
mssymb,floatfix]{revtex4}
\usepackage{graphics}
\usepackage{graphicx}% Include figure files
\usepackage{bm}% bold math

\begin{document}

\newcommand*{\bra}[1]{\ensuremath{\left\langle #1 \right\vert }}

\newcommand*{\ket}[1]{\ensuremath{ \left\vert #1 \right\rangle }}

\newcommand*{\braket}[2]{\ensuremath{  \left\langle #1 | #2 \right\rangle}}

\newcommand*{\E}[1]{\ensuremath{  \left\langle #1 \right\rangle}}

\newcommand*{\partiald}[2][]{\ensuremath{  \frac{\partial #1}{\partial #2} }}

\newcommand*{\totald}[2][]{\ensuremath{  \frac{d#1}{d#2} }}

\newcommand*{\vect}[1]{\ensuremath{  \mathbf{#1}}}

\newcommand{\displacement}[2][]{\ensuremath{  \hat D^{#1} \negmedspace \left( \left\{ #2 \right\}\!\right)}}

\newcommand*{\sbra}[1]{\ensuremath{ \left( #1 \right\vert}}

\newcommand*{\sket}[1]{\ensuremath{ \left\vert #1 \right) }}

\newcommand*{\sbraket}[2]{\ensuremath{ \left( #1 | #2 \right)}}
 
\newcommand*{\trace}[1]{\ensuremath{\text{Tr} \left[ #1 \right]}}

\title{Quantum state reconstruction via continuous measurement}

\author{Andrew Silberfarb}
\email{drews@unm.edu}
\affiliation{Department of Physics and Astronomy, University of New Mexico, Albuquerque, NM 87131}
 
\author{Poul S. Jessen}
\affiliation{Optical Sciences Center, University of Arizona, Tucson Arizona 85721}

\author{Ivan H. Deutsch}
\affiliation{Department of Physics and Astronomy, University of New Mexico, Albuquerque, NM 87131}

\date{\today}

\begin{abstract}
We present a new procedure for quantum state reconstruction based on weak continuous 
measurement of an ensemble average. By applying controlled evolution to the initial 
state new information is continually mapped onto the measured observable.  A 
Bayesian filter is then used to update the state-estimate in accordance with the measurement 
record. This generalizes the standard paradigm for quantum tomography based on strong, 
destructive measurements on separate ensembles.  This approach to state estimation can 
be non-destructive and real-time, giving information about observables whose evolution 
cannot be described classically, opening the door to new types of quantum feedback control. 
\end{abstract}


\maketitle

The control of quantum mechanical systems is finding new applications in information 
processing tasks such as cryptography and computation \cite{Nielsen2000}.  
Experimental reconstruction of a quantum state is thus essential to verify preparation, to 
detect the presence of errors due to noise and decoherence, and to determine the fidelity 
of control protocols using process tomography.  Moreover, real-time ``state estimation'' 
may allow improvement of precision metrology beyond the standard quantum limit 
\cite{Geremia03}, with the possibility of active control through closed-loop feedback 
protocols \cite{Geremia04}.  In addition, measurement of the quantum state can provide 
information about unknown or nontrivial dynamics, such as those arising in the study of 
quantum chaos.  Laboratory demonstrations of state reconstruction are numerous and 
span a broad range of physical systems, including light fields \cite{smithey93}, 
molecules \cite{dunn95}, ions  \cite{leibfried96}, atoms \cite{kurtsiefe97}, spins 
\cite{chuang98, Klose01}, and entangled photon pairs \cite{white99}.

In this letter we consider a new protocol for quantum state reconstruction based on 
continuous, weak measurement of a single observable on a single ensemble of identically 
prepared systems.  The ensemble is driven so that each member undergoes an identical, 
carefully designed dynamical evolution that continually maps new information onto the 
measured quantity.  This is in contrast to the standard paradigm for quantum state 
reconstruction based on strong and therefore destructive measurements, often of a large 
set of observables performed on many copies of the unknown state.  Our weak 
measurement approach has a number of possible advantages in situations that lend 
themselves naturally to working with ensembles.  Strong measurements on ensembles are 
inefficient because only a single observable can be measured after each preparation and 
the information gained about the observable is extracted independent of the required 
fidelity. By contrast, weak measurements can be optimized to obtain just enough 
information to estimate the density matrix to some required fidelity, in real time, and with 
minimal disturbance of each member. One can then imagine using the extracted 
information to perform closed-loop feedback control based on knowledge about the entire 
quantum state, rather than on conventional ``state estimation" of Gaussian random 
variables that evolve according to classical dynamics, as considered to date 
\cite{Molmer04}.  Our procedure is broadly applicable in systems where continuous 
weak measurement tools have been developed, such as nuclear magnetic resonance in 
molecules \cite{chuang98} and polarization spectroscopy in dilute atomic vapors 
\cite{Smith03}, but where noise and decoherence limits the ability to perform strong 
measurements regardless of the amount of signal averaging. 

To perform quantum state reconstruction one considers a set of measurements   $\{ 
M^{(i) }\}$, each of which has a set of outcomes  $m_j^{(i)}$,  
$\{j=1,2,\ldots,j^{(i)}_{\text{max}}\}$.  The set is said to be ``informationally complete" 
if for a given state $\rho_0$  the set of probabilities $\{ p_j^{(i)} \}$  of the measurement 
outcomes can be inverted to determine $\rho_0$.  In our protocol the probabilities are 
assigned using a single ensemble  $\rho_N =  \rho_0^{\otimes N}$, whose dynamical 
evolution is driven in a known fashion and monitored by a probe that measures the sum 
of the identical observables  $\{O\}$  on each member.  Due to the central limit theorem 
the measurement record of this probe has the form
\begin{equation} \label{E:Measurement}
M(t)= N \E{O}_t  + \Delta M(t),
\end{equation}
where $\E{O}_t$ is the quantum expectation value at time $t$, and $\Delta M(t)$ is a 
Gaussian white noise process with variance  $\sigma^2 = 1/\kappa \Delta t$ for 
measurement strength $\kappa$ and detector averaging time $\Delta t$.  In principle a 
measurement of the collective observable $N\E{O}_t$ leads to backaction on the 
collective many-body state and can cause individual members of the ensemble to become 
correlated \cite{Kuzmich00,Geremia04}. Such correlations influence the outcome of 
future measurements and greatly complicate the task of reconstructing the initial state 
$\rho_0$.  Additionally, the gain from performing such quantum limited measurements is 
small, as the majority of the information about the state of individual ensemble members 
has already been extracted by the probe prior to reaching the quantum limited regime. We 
thus restrict our considerations to cases where the measurement uncertainty, averaged 
over the total measurement time $T$, is large compared to the intrinsic quantum 
uncertainty (projection noise) of the collective observable, $1/\kappa T > N \Delta O^2$, 
and backaction onto the collective state is insignificant.  Experimentally this is also the 
most common situation.  Of course a sufficient measurement signal-to-noise ratio must 
still be available to reconstruct the state of an individual member of the ensemble.  This 
requires $N>>1$ so that the quantum backaction associated with information gain is 
distributed uniformly among the entire ensemble, with negligible disturbance of any 
single member state. 

The goal is to invert the measurement history, Eq.\ (\ref{E:Measurement}), to determine 
$\rho_0$.   As we wish this procedure to be independent of $\rho_0$, it is most 
convenient to work in the Heisenberg picture and express $\E{O}_t=\trace{O(t) \rho_0 } 
= \sbraket{O(t)}{\rho_0}$, where in the second equality we have written the trace as an 
inner-product between ``superoperators" \cite{Caves99}.  We coarse grain over the 
detector response time $\Delta t$, such that $\sbra{O_i} = \int_t^{t+dt} \sbra{O(t)} 
dt/\Delta t$, obtaining a discrete measurement history time-series $\{M_i\}$, with $M_i 
=  N \sbraket{O_i}{\rho_0} +  \sigma W$, where now the measurement operators 
$\{O_i\}$ are determined in advance by the known dynamics, and where $W$ is a 
Gaussian random variable with zero mean and unit variance.  This equation recasts the 
reconstruction problem as a stochastic linear estimation problem for the underlying state 
$\rho_0$.

In order to reconstruct the state from the measurement time-series, the set of 
measurements operators $\{ O_i \}$ must be informationally complete, spanning the 
space of density operators, i.e., the dynamics must map the initial measurement to all 
possible (Hermitian) measurements. This is best achieved by introducing an explicit set 
of control parameters, with a time dependent series of Hamiltonians $\{H_i\}$.  We 
require that this set generate the Lie algebra for $SU(d)$, where $d$ is the dimension of 
the space; the system must be ``controllable''.  Generally, the evolution will have 
some associated decoherence that will degrade the measurement.  Including these terms, 
the evolving measurement operators can then be expressed in terms of the base 
observable as $\sbra{O(t)} = \sbra{O} \mathcal{S}_t$ where  $\mathcal{S}_t  =  
\mathbb{T} exp \left[ \int_0^{t} dt' \,\mathcal{L}_{t'} \right]$  with $\mathcal{L}_t$ the 
generator of the dynamics, and $\mathbb{T}$ the time ordering operator.  To simulate 
this evolution we consider a time scale $\delta t$ over which $\mathcal{L}_t$ changes 
negligibly. The semigroup property then allows us to approximate 
$\mathcal{S}_{t+\delta t} = e^{i \mathcal{L}_t \delta t} \mathcal{S}_t$ which can be 
numerically calculated given a system of reasonable size, $d < 100$. 

A Bayesian filter determines how our knowledge of $\rho_0$ is updated due to a 
measurement history $\{M_i\}$, 
\begin{equation}
	P(\rho_0|\{M_i\}) = A P(\{M_i\}|\rho_0) P(\rho_0).
\end{equation}
Here $A$ is the normalization constant for the posterior distribution and $P(\rho_0)$ 
contains the prior information, including the fact that $\rho_0$ is a valid density matrix 
(ie. has trace one and is positive).  $P(\{M_i\}|\rho_0)$ is the conditional distribution, 
which contains the information gained during the experiment.  The conditional 
distribution thus quantifies how well the measurement performs.  Due to the Gaussian 
measurement statistics, this distribution has the form,
\begin{equation}\label{E:condin}
	P(\{M_i\}|\rho_0) \propto  \exp \left[ - \sbra{\delta \rho} \mathcal{R} \sket{\delta 
\rho}  \right].
\end{equation}
The superoperator $\mathcal{R}$ is the covariance matrix for the measurements and 
$\delta \rho=\rho_0-\hat{\rho}$ is the difference between the prepared state and the 
maximum likelihood estimate of this state given the measurements, equivalent to the least 
squares estimator for a Gaussian random variable,
\begin{equation}\label{E:cov}
	\mathcal{R} = \frac{1}{\sigma^2} \sum_i \sket{O_i}\sbra{O_i}, \quad
	\sket{\hat{\rho}} = \frac{1}{\sigma^2} \sum_i M_i \mathcal{R}^{-1} \sket{O_i} 
. 
\end{equation}
This evolving covariance matrix generalizes the classical update rule discussed in 
\cite{Molmer04}.  For systems beyond spin-1/2, full state reconstruction requires 
information about higher moments of observables, whose evolution is fundamentally 
quantum mechanical.  The conditional probability distribution has entropy
\begin{equation}\label{E:entropy}
	S = - \log \mathcal{R} = - \sum_j \log \lambda_j,
\end{equation}
where $\lambda_j$ are the eigenvalues of the covariance matrix, corresponding to the 
inverse of the variances of Eq.\ (\ref{E:condin}) along its primary axes.  Qualitatively 
$\sqrt{\lambda_j}$ is the signal-to-noise-ratio with which we can extract a measure of 
one of the eigen-operators of  $\mathcal{R}$ from the measurement record. This entropy 
thus provides a collective measure of the information gained about all parameters, 
independent of the initial state and any prior information.  To obtain an accurate 
reconstruction we need to optimize the entropy and any additional costs over the free 
parameters (controls).

Given the measured information, one estimates the quantum state using the mean of the 
Gaussian conditional distribution, as given by the least squares fit, Eq.\ (\ref{E:cov}).  
Because this does not take into account the prior information, the estimated state 
$\hat{\rho}$ is not automatically a valid density matrix.  To correct this, we must enforce 
$\trace{\hat{\rho}} = 1$ and positivity.  The trace condition is ensured by adding a 
pseudo-measurement of the trace $M_0 = I/d$ which has variance $\sigma_0 = 0$.  To 
enforce positivity one could solve for the closest positive state to $\hat{\rho}$ using 
convex optimization \cite{Vandenberghe96}.  Alternatively one can get a reasonable and 
much simpler estimate by setting the negative eigenvalues of $\hat{\rho}$ to zero, and 
renormalizing to give $\hat{\rho}_\text{pos}$, sufficient when the estimated state has 
good fidelity.   This procedure is used in the example below.

As a concrete demonstration of the power and versatility of our method we consider the 
reconstruction of the quantum state associated with the total spin-angular momentum of 
an ensemble of alkali atoms, in our specific example the $F=3$ or $F=4$ hyperfine 
manifolds of the $6S_{1/2}$ ground state of $^{133}\text{Cs}$. The number of 
parameters needed for reconstruction are then $(2F+1)^2 - 1$, giving $48$ and $80$ 
components respectively.  Consider a cloud of cesium atoms prepared in identical states 
$\rho_0$ and coupled to a common, linearly polarized probe beam tuned near the D1 
($6S_{1/2} \to P_{1/2}$) or D2 ($6S_{1/2} \to P_{3/2}$) resonance \cite{Smith03}. 
Information about the atomic spins is obtained by measuring the Faraday rotation of the 
probe polarization, which provides a continuous measurement of the spin component 
along the direction of propagation, $O = F_z$.  Shot noise in the probe polarimeter gives 
rise to the fluctuations $W$ which limit the measurement accuracy.  

In the regime of strong backaction onto the collective spin state, such measurements have 
been used to generate spin squeezed states \cite{Kuzmich00,Geremia04}, and to perform 
sub-shot noise magnetometry \cite{Geremia04,Molmer04}.  In the regime of negligible 
backaction that is of interest here, Smith {\em et al.} continuously monitored the Larmor 
precession of spin in an external magnetic field, and observed a series of dynamical 
collapse and revivals due to a nonlinear term in the spin Hamiltonian \cite{Smith04}.  
While this nonlinear collapse limits the observation window of a quantum nondemolition 
measurement, it also allows for full controllability of the atomic spin so that in principle 
one can reconstruct the input quantum state according to the procedure described above. 
The nonlinearity results from the AC Stark shift caused by off-resonance excitation of the 
D1 or D2 transition, and is directly proportional to the excited state hyperfine splitting.  
Off-resonance excitation also introduces a small but unavoidable amount of decoherence 
due to photon scattering. Quantum state reconstruction requires a large enough 
nonlinearity to generate dynamics that cover the entire operator space before decoherence 
erases information about the initial state.  This, therefore, favors large excited state 
hyperfine splittings.

To control the system we apply a time dependent magnetic field.  The overall 
Hamiltonian, including the nonlinear AC Stark shift induced by an $x$-polarized probe, 
is \cite{Smith04}
\begin{equation}\label{E:Hamiltonian}
 H(t) = g_F \mu_B  \left( \mathbf{B}(t) + \mathbf{B}_{0} \right) \cdot \mathbf{F} +  
\beta \hbar \gamma F_x^2
\end{equation}
where $\mathbf{B}(t)$ is the control field and $\mathbf{B}_{0}$ represents any 
background field that might be present in an experiment. In the nonlinear term we have 
factored out the scattering rate $\gamma$ for a transition with unit oscillator strength, and 
introduced the ratio $\beta$ between the timescales for coherent evolution and 
decoherence due to optical pumping. In this example we explore two regimes: a probe 
detuned from the D2 transition by much more than the excited state hyperfine splitting 
($\beta=0.81$), and a probe tuned halfway between the two excited hyperfine states 
($\beta=7.67$). Note that the interaction with the magnetic field alone only generates 
$SU(2)$ rotations which, for $F>1/2$,  is insufficient to generate the full $SU(2F+1)$ 
algebra.  The evolution of the ensemble is governed by the master equation $\mathcal{L}_t [\rho ] = - \frac{i}{\hbar} [H(t),\rho] - \frac{\gamma}{2} 
\mathcal{D}[\rho],$ where $\rho$ has support only on the ground state of interest and all other states have 
been adiabatically eliminated \cite{Cohen1992}.  The superoperator $\mathcal{D}[\rho]$ 
includes optical pumping within and out of the initial hyperfine manifold. To simulate 
this evolution we construct the $\mathcal{L}_t$ for some choice of scattering rate, 
background field, and controls $\mathbf{B}(t)$.  The measurement strength $\kappa$ is 
determined empirically by the shot-noise limited measurement uncertainty $\sigma$, 
which we characterize by the signal-to-noise-ratio $SNR = M_\text{max}/\sigma$, 
where $M_\text{max} = \max_{\rho} \sbraket{F_z}{\rho}$ is the maximum signal 
possible. We account for inhomogeneous values of  $\mathbf{B}_{0}$ by averaging 
over a Gaussian distribution corresponding to a standard deviation of $60$ Hz in the 
induced Larmor frequency.  The duration of the simulated measurement is $T = 4$ms, 
the coarse graining time is $4 \mu \text{s}$, and the average photon scattering rate is 
$\gamma = 10^3 \text{s}^{-1}$.  Finally we simulate the effect of a low pass filter by 
averaging our measurement over a few coarse graining time steps.

\begin{figure}
	\scalebox{.5}{\includegraphics{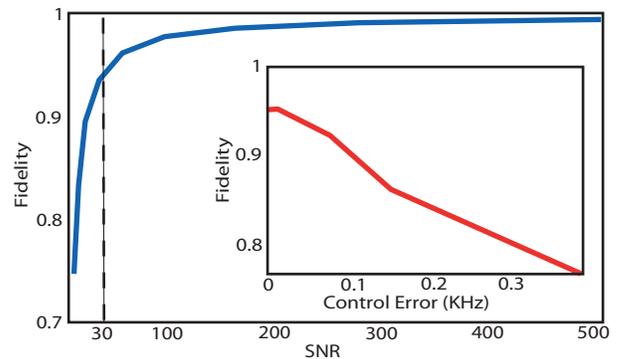}}
	\caption{Quantum state reconstruction of the "cat" state described in the text, for 
the $F=3$ hyperfine ground state of $^{133}\text{Cs}$ probed on the D1 transition. The 
figure shows the fidelity of reconstruction as a function of the measurement signal-to-
noise ratio. Large nonlinearity and little decoherence allows good fidelity reconstruction 
in a single run, as indicated by the dashed line where a fidelity $\mathcal{F}=0.95$ is reached 
already at a very modest SNR  of 30.  The inset shows how the reconstruction degrades 
due to errors in the control field, at a constant SNR of 30.  For a reasonable $1\%$ 
control field error the fidelity drops to $\mathcal{F}=0.85$.\label{F:D1}}
\end{figure}

The dynamical generation of a complete set of measurements incurs an unavoidable 
decoherence cost and so must be done as efficiently as possible.  To this end we optimize 
based on a cost function consisting of the entropy $S(\mathcal{R})$ (Eq.\ 
\ref{E:entropy}) and any additional control costs $C(\mathbf{B})$.  In this case the 
additional costs include  the degradation in reconstruction due to loss of field control at 
large amplitudes, and inability to rapidly change the fields.   We minimize these costs 
over all possible time dependent magnetic fields $\mathbf{B}(t)$.  This is done 
sequentially, first restricting the magnetic fields to optimize the control costs 
$C(\mathbf{B})$, and then optimizing the entropy subject to these restrictions.  The 
magnetic fields is restricted to be in the $x-y$ plane, with magnitude $|\mathbf{B}| = B$, 
such that the Larmor frequency is $\omega_B = \mu_B B/h = 15$ kHz.  Additionally we 
specify the field at only $n=50$ independent times and smoothly interpolate between 
them to ensure slow variation.  The only free parameters are thus a set of $n=50$ 
indepedent angles.  Optimizing the entropy subject to these constraints, we find that the 
landscape has many local minima, precluding the use of purely local search techniques.  
Instead, we use a one dimensional global search, where we iteratively optimize one of the 
$n = 50$ independent angles, holding the others fixed.  This process is repeated until all 
of the angles are globally stationary, within some tolerance, assuming the others are held 
fixed.  This procedure is suboptimal, but converges reasonably well.

\begin{figure}
	\scalebox{.83 }{\includegraphics{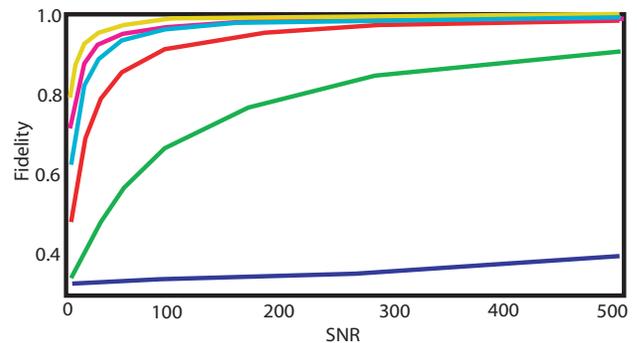}}
	\caption{ Quantum state reconstruction for the $F=4$ hyperfine state probed on 
the D2 transition, where the nonlinearity is weaker than for the case in Fig.\ \protect{\ref{F:D1}}.  
The curves correspond to an increasing number of independent runs, which allow 
increasingly complete exploration of the operator space.  From lowest to highest fidelity 
we use: 1 run (blue), 2 runs (green), 3 runs (red), 4 runs (cyan), 5 runs (magenta), and 6 
runs (yellow).\label{F:D2}}
\end{figure}

The results of simulated reconstructions are shown in Figs.\ \ref{F:D1} and \ref{F:D2}.  
Given an initial preparation in the $F=3$ hyperfine manifold, in the ``cat state" 
$\ket{\psi_0}= (\ket{m= 3} + \ket{m = -3})/\sqrt{2}$, the fidelity of the reconstruction, 
$\mathcal{F}= \bra{\psi_0} \hat{\rho}_\text{pos}\ket{\psi_0}$, averaged over 1000 
noise realizations, is plotted versus simulated SNR.  Increasing SNR clearly results in a 
better reconstruction.  The parameters needed to attain good fidelity for reconstruction of 
the $F=3$ ground state using the D1 transition (Fig.\ \ref{F:D1}) appear to be well within 
the reach of current experiments at $\mathcal{F}\approx .95, \text{SNR}=30$ \cite{Smith03}.  
Even with a possible $1\%$ uncertainty in the control fields, a fidelity of $\mathcal{F} = .85$ should 
be possible (Fig.\ \ref{F:D1} inset).  A different regime is illustrated by reconstruction of 
a spin $F=4$ using the D2 resonance.  This turns out to be infeasible with a single 
measurement run performed on a single ensemble, even assuming very large SNR (Fig.\ 
\ref{F:D2}). This is because more parameters must be reconstructed, $80$ versus $48$, 
and because the decoherence incurred in conjunction with the nonlinearity is larger due to 
the smaller hyperfine splitting of the $P_{3/2}$ manifold. It is still possible, however, to 
obtain a high fidelity reconstruction if we combine the measurement records from 
multiple independent runs that each start with a fresh ensemble and explore operator 
space in different ways.

We have presented a new protocol for quantum state reconstruction based on continuous 
measurement of an ensemble of $N$ members and demonstrated our procedure through a 
simulated reconstruction of a spin $J$ via polarization spectroscopy of a gas of cold 
atoms.  The reconstruction technique is nondestructive and exploits classical estimation 
theory, providing a starting point for consideration of more complex applications of 
quantum control tasks such as quantum feedback.  In the future work we plan to improve 
our optimization procedure for robustness in control-parameter uncertainty and examine 
global search procedures such as convex optimization  \cite{Vandenberghe96}.  The 
tools developed here should provide new avenues for real-time state quantum estimation 
that allow us to explore the dynamical generation of nonclassical features, such as 
entanglement.  This is of particular interest for mesoscopic systems whose classical 
description exhibits chaos \cite{Lakshminarayan01}.
\acknowledgments{
AS would like to thank Greg Smith and Souma Chaudhury for useful discussion.  IHD 
and AS were supported by the NSF grant PHY-0355040 and NSA/ARDA grant 
DAAD19-01-1-0648.  PSJ was supported by NSF grant PHY-0099528 and 
DARPA/ARO grant DAAD 19-01-1-0589.
}

\end{document}